\documentclass[referee,pdflatex]{sn-jnl}
\usepackage{graphicx} 
\usepackage{amsmath}
\usepackage{bm}

\usepackage{times}

\newcommand{\subs}[1]{$_{\mathrm {#1}}$}
\newcommand{\subm}[1]{_{\mathrm {#1}}}

\renewcommand{\deg}{^{\circ}}
\newcommand{\Tc}{T\subm{c}}

\newcommand{\Hcc}{H\subm{c2}}

\newcommand{\vF}{\bm{v}\subm{F}}

\newcommand{\BS}{Bi$_2$Se$_3$}

\newcommand{\bs}{Bi$_2$Se$_3$}
\newcommand{\csvs}{CsV$_3$Sb$_5$}
\newcommand{\rbvs}{RbV$_3$Sb$_5$}
\newcommand{\kvs}{KV$_3$Sb$_5$}
\newcommand{\Avs}{$A$V$_3$Sb$_5$}

\newcommand{\Hsix}{H^{(6)}_{\mathrm{c2}}}
\newcommand{\Htwo}{H^{(2)}_{\mathrm{c2}}}
\newcommand{\Hzero}{H^{(0)}_{\mathrm{c2}}}

\title[\ ]{Violation of Emergent Rotational Symmetry in the Hexagonal Kagome Superconductor CsV$_{\bm{3}}$Sb$_{\bm{5}}$}

\author[1]{\fnm{Kazumi} \sur{Fukushima}}
\author[1]{\fnm{Keito} \sur{Obata}}
\author[1]{\fnm{Soichiro} \sur{Yamane}}
\author[1]{\fnm{Yajian} \sur{Hu}}
\author[2,3,4]{\fnm{Yongkai} \sur{Li}}
\author[2,3]{\fnm{Yugui} \sur{Yao}}
\author*[2,3,4]{\fnm{Zhiwei} \sur{Wang}}
\email{zhiweiwang@bit.edu.cn}
\author[1,5]{\fnm{Yoshiteru} \sur{Maeno}}
\author*[1]{\fnm{Shingo} \sur{Yonezawa}}\email{yonezawa.shingo.3m@kyoto-u.ac.jp}

\affil[1]{\orgdiv{Department of Physics, Graduate School of Science}, \orgname{Kyoto University}, \orgaddress{\city{Kyoto}, \postcode{606-8502}, \country{Japan}}}

\affil[2]{\orgdiv{Key Laboratory of Advanced Optoelectronic Quantum Architecture and Measurement, Ministry of Education (MOE), School of Physics}, \orgname{Beijing Institute of Technology}, \orgaddress{\city{Beijing}, \postcode{100081}, \country{P. R. China.}}}

\affil[3]{\orgdiv{Beijing Key Lab of Nanophotonics and Ultrafine Optoelectronic Systems}, \orgname{Beijing Institute of Technology}, \orgaddress{\city{Beijing}, \postcode{100081}, \country{P. R. China.}}}

\affil[4]{\orgdiv{Material Science Center,  Yangtze Delta Region Academy}, \orgname{Beijing Institute of Technology}, \orgaddress{\city{Jiaxing}, \postcode{314011}, \country{P. R. China.}}}

\affil[5]{\orgdiv{Toyota Riken-Kyoto University Research Center (TRiKUC)}, \orgname{Kyoto University}, \orgaddress{\city{Kyoto}, \postcode{606-8501}, \country{Japan}}}

\abstract{
Superconductivity is caused by electron pairs that are canonically isotropic, whereas some exotic superconductors are known to exhibit non-trivial anisotropy stemming from unconventional pairings.
However, superconductors with hexagonal symmetry, the highest rotational symmetry allowed in crystals, exceptionally have strong constraint that is called emergent rotational symmetry (ERS): anisotropic properties should be very weak especially near the critical temperature $\Tc$ even for unconventional pairings such as $d$-wave states.
Here, we investigate superconducting anisotropy of the recently-found hexagonal Kagome superconductor \csvs, which is known to exhibit various intriguing phenomena originating from its undistorted Kagome lattice formed by vanadium atoms.
Based on calorimetry performed under accurate two-axis field-direction control, we discover a combination of six- and two-fold anisotropies in the in-plane upper critical field.
Both anisotropies, robust up to very close to $\Tc$, are beyond predictions of standard theories.
We infer that this clear ERS violation with nematicity is best explained by multi-component nematic superconducting order parameter in \csvs\ intertwined with symmetry breakings caused by the underlying charge-density-wave order.
}

\begin{document}

\maketitle

\pagestyle{myheadings}
\markboth{}{}


\begin{center}\textit{\today}\end{center}

\section*{Anisotropies in superconductors}

Superconductivity occurs as a consequence of formation of electron pairs called the Cooper pairs.
In standard microscopic theories Cooper pairs are assumed to be isotropic~\cite{Bardeen1957.Phys.Rev.108.1175}. 
Similarly, standard theories for macro- or meso-scopic properties, the Ginzburg-Landau (GL) formalisms, also canonically assumes isotropic  macroscopic quantum-mechanical wavefunction as the superconducting (SC) order parameter~\cite{Ginzburg1950.SovPhysJETP.20.1064}.
Experimentally, indeed, most ordinary superconductors exhibit rather isotropic SC properties or, at most, trivial anisotropy inherited from normal-state electronic properties~\cite{TinkhamText2}.
However, some exotic superconductors exhibit non-trivial anisotropic properties stemming from order-parameter structures, and
both microscopic as well as phenomenological theories have been extended to deal with such unconventional superconductivity.
A well-known example is the $d$-wave SC state having sign changes in the order parameter upon 90-degree rotation, hosting various interesting anisotropic properties~\cite{Tsuei2000.RevModPhys.72.969,Sakakibara2016.RepProgPhys.79.094002}. 
Another exotic example is nematic superconductivity~\cite{Fu2014.PhysRevB.90.100509}, which is recently identified first in doped \BS\ superconductors~\cite{Matano2016.NaturePhys.12.852,Yonezawa2017.NaturePhys.13.123,Pan2016.SciRep.6.28632,Yonezawa2019.condmat.4.2} and subsequently in other systems such as bilayer graphene~\cite{Cao2021.Science.372.6539}.
Nematic superconductors exhibit two-fold anisotropy in bulk SC properties originating from rotational-symmetry breaking in the order-parameter amplitude.
Finding a new species of superconductivity with unique anisotropy is one of the most important goals of fundamental research of superconductivity.

Hexagonal materials are exceptionally unique when in the study of novel superconductivity. 
On one hand, the six-fold rotational symmetry, the highest rotational symmetry allowed in crystals, enables the system to realize interesting SC order parameters.
In particular, various two-component superconducting order parameters, such as chiral or nematic $p$, $d$, or $f$-wave superconductivity, are allowed owing to the degeneracy between $(x,y)$ or $(xy, x^2-y^2)$ basis functions~\cite{Sigrist1991.RevModPhys.63.239}.
On the other hand, hexagonal systems have a hidden special feature: six-fold rotational symmetry is equivalent to cylindrical symmetry up to the fourth order terms in the GL theory~\cite{Burlachkov1985.SovPhysJETP.62.800,Agterberg1995.PhysRevB.51.8481,Sauls1996.PhysRevB.53.8543}. 
Thus, for example, the upper critical field $\Hcc$ in the plane perpendicular to the six-fold axis should be perfectly isotropic.
This isotropy can be removed by considering sixth-order terms, allowing the presence of weak six-fold $\Hcc$ anisotropy. 
Nevertheless, such in-plane hexagonal $\Hcc$ anisotropy, $\Hsix$, if exists, should rapidly vanish for $T \to \Tc$ as $\Hsix \propto (1-T/\Tc)^3$~\cite{Sauls1996.PhysRevB.53.8543,Krotkov2002.PhysRevB.65.224506}.
This phenomenon is called the emergent rotational symmetry (ESR) in hexagonal superconductors~\cite{Venderbos2016.PhysRevB.94.094522}.
ESR violation, namely six-fold anisotropy that is robust with increasing temperature, is fully non-trivial. 
Experimentally, in-plane anisotropy in various hexagonal superconductors have been tested~\cite{Stokan1979.PhysRevB.20.3670,Keller1994.PhysRevLett.73.2364,Ghosh2003.PhysRevB.68.054507,Shi2003.PhysRevB.68.104513,Zuo2017.PhysRevB.95.014502,Yasuzuka2019.JSupercondNovMagn.33.953}, but hexagonal anisotropy has only been reported in a very few examples using transport measurements~\cite{Stokan1979.PhysRevB.20.3670,Keller1994.PhysRevLett.73.2364,Zuo2017.PhysRevB.95.014502}; and it has never been reported using thermodynamic probes.
Moreover, detailed temperature dependence of hexagonal anisotropy has never been elucidated.

Here, in this Article, we report discovery of combined hexagonal and nematic SC properties in the hexagonal material \csvs. 
Our high-resolution specific-heat measurements performed under precise two-axis control of magnetic field directions reveal six- and two-fold anisotropies in bulk $\Hcc$ when the field is rotated in the hexagonal $ab$ plane.
Our results marks the first thermodynamic evidence for the violation of the ERS in hexagonal superconductors, indicating realization of SC states beyond the ordinary GL formalism.
We argue that our result is best explained by two-component order parameter coupled to the underlying charge-density-wave (CDW) order, which is believed to weakly break the nominal hexagonal symmetry.

\section*{Kagome superconductor \csvs}

Our target material \csvs\ along with its sister compounds \kvs\ and \rbvs\ have been extensively studied recently because of their fascinating properties originating from undistorted Kagome net of vanadium, unconventional CDW order, and superconductivity~\cite{Ortiz2019.PhysRevMaterials.3.094407,Ortiz2020.PhysRevLett.125.247002}. 
In \csvs, the CDW occurs below 94~K followed by superconductivity below $\Tc \sim 3$~K. 
Both the CDW and SC states are of utmost interest, and the symmetry properties in these orders have been widely debated~\cite{Neupert2022.NaturePhys.18.137}.
The CDW is now believed to be a bond order in the Kagome lattice, resulting in the so-called star-of-David (SoD) or tri-hexagon (TrH) deformation~\cite{Ortiz2021.PhysRevX.11.041030,Tazai2022.SciAdv.8.eabl4108}. 
The CDW also has four-unit-cell $c$-axis modulation and thus it is characterized by a $2\times 2\times 4$ expansion of the unit cell~\cite{Ortiz2021.PhysRevX.11.041030}.
The detailed crystal structure in the CDW phase is still under debate. 
A recent study suggest a trigonal model with the overall $P\bar{3}$ space group~\cite{Ortiz2021.PhysRevX.11.041030}.
Additional time-reversal-symmetry breaking~\cite{Yu2021.arxiv.2107.10714} and nematicity (rotational symmetry breaking) has also been reported~\cite{Xiang2021.NatureCommun.12.6727,Nie2022.Nature.604.59}.
Nevertheless, even in the CDW phase, the overall lattice distortion as well as accompanied changes in the electronic state is rather subtle~\cite{Ortiz2021.PhysRevX.11.041030,Nakayama2021.PhysRevB.104.L161112,Kang2022.NaturePhys.18.301}, and the hexagonal symmetry remains as a good starting point.

Reflecting the underlying symmetry breakings in this compound, the SC state can have highly unconventional nature.
However, the fundamental information of such as gap structure and/or order-parameter symmetry has not been established.  
Several reports claim fully-gapped conventional SC properties but with strong multi-band nature~\cite{Mu2021.ChinPhysLett.38.077402,Duan2021.SciChinaPhysMechAstron.64.107462,Roppongi2022.arxiv.2206.02580}, whereas nodal properties have been suggested from thermal transport and scanning tunneling microscopy studies~\cite{Xu2021.PhysRevLett.127.187004,Zhao2021.arxiv.2102.08356,Chen2021.Nature.599.222}.
Moreover, SC order parameter is recently found to accompany atomic-scale modulation, which is called the roton pair-density-wave (PDW) state~\cite{Chen2021.Nature.599.222}.
To reveal the true nature of the superconductivity, study of in-plane anisotropy is inevitably important because it reflects various properties of the SC order parameter~\cite{Machida1985.JPhysSocJpn.54.1552}. 
For \csvs, several transport studies have reported two-fold in-plane anisotropy in SC properties~\cite{Xiang2021.NatureCommun.12.6727,Ni2021.ChinPhysLett.38.057403}, suggesting nematic feature in the SC state.
However, for quasi-two-dimensional (Q2D) materials such as \csvs, very accurate field alignment using two-axis field control is necessary to avoid any extrinsic features originating from field misalignment.
To our knowledge, in-plane anisotropy studies performed under two-axis field-direction control have not been reported to date.

\section*{Calorimetry using a high-quality sample}

\begin{figure}[h]
\begin{center}
\includegraphics[width=\linewidth]{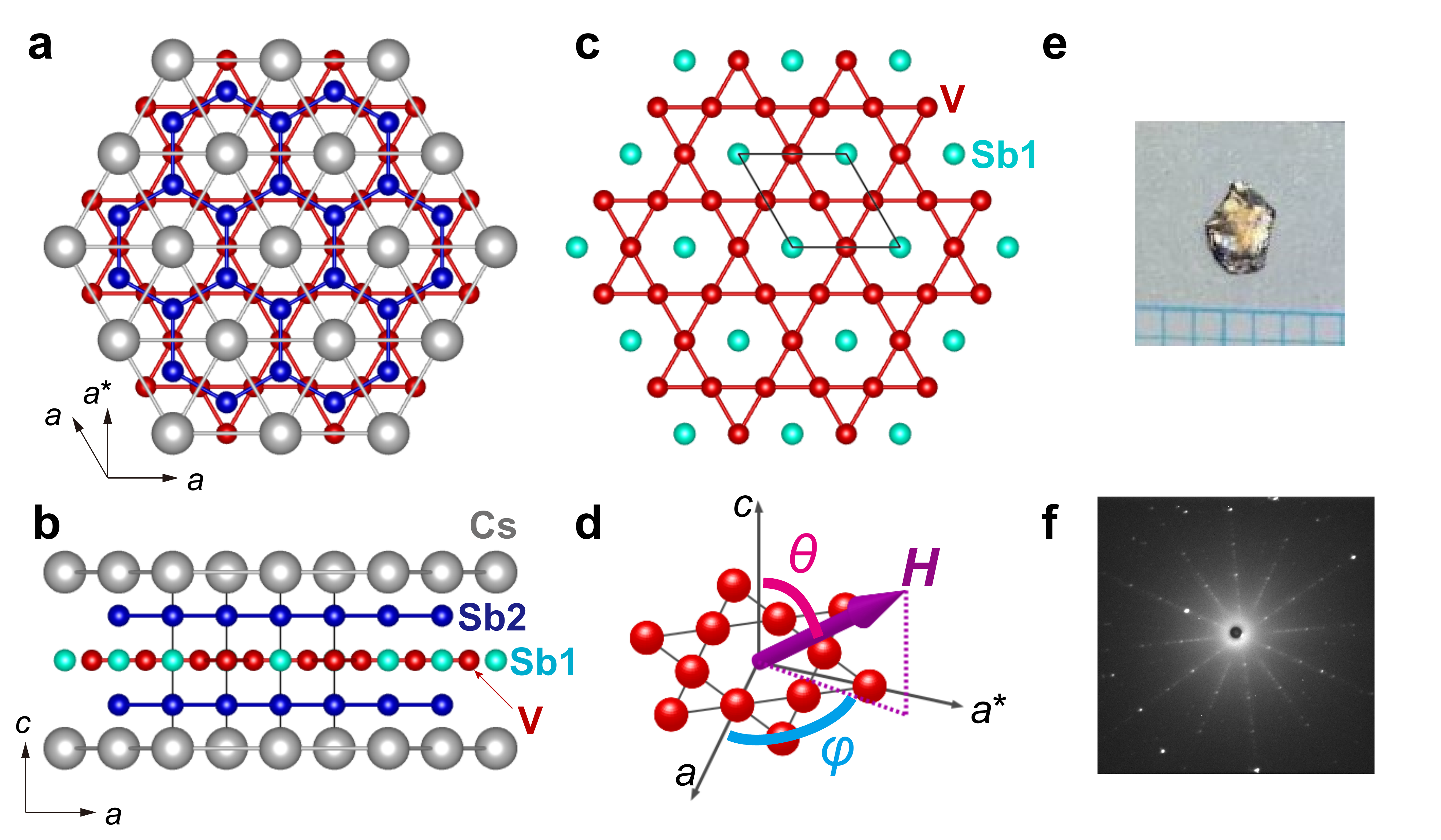}
\end{center}
\caption{
{\bf Hexagonal crystal structure of the Kagome superconductor \csvs.}
{\bf a-b}. Crystal structure viewed from the $c$ axis ({\bf a.}) and from the $a^\ast$ axis ({\bf b.}). 
Cs and V are represented by the gray and red spheres. 
Sb are shown with light-blue (Sb1 site) and blue (Sb2 site) spheres.
{\bf c}. Schematic of the vanadium Kagome layer. The Kagome net of V is formed with Sb1 located at the center of hexagons.
{\bf d}. Definition of the magnetic field angles $\theta$ and $\phi$.
{\bf e}. Photo of the sample. The size of this sample is 3 $\times$ 2 $\times$ 0.2~mm$^3$. The mass is 3.8921~mg. 
{\bf f}. Laue photo of the sample. Clear Laue spots indicate high crystallinity of the sample.
\label{fig1}
}
\end{figure}

In this study, we investigated in-plane and out-of-plane anisotropies of the superconductivity in \csvs\ by means of field-angle-resolved calorimetry.
Specific heat $C$ was measured with a hand-made calorimeter based on the AC method.
We used a high-quality single crystal grown by a self-flux method (Fig.~\ref{fig1}{\bf e}). 
Laue photos of this sample shows clear spots as shown in Fig.~\ref{fig1}{\bf f}. 
We found that some other samples exhibits much broader spots, presumably due to stacking faults. 
Thus, the clear Laue spots in our sample evidences high crystalline quality.
Moreover, we have noticed that samples with broader Laue spots tend to exhibit broader CDW or SC transitions, indicaing importance of choosing samples with high crystallinity.

Figure.~\ref{fig2}{\bf a} shows magnetic susceptibility of this sample and
Fig.~\ref{fig2}{\bf{b}} the temperature $T$ dependence of the electronic specific heat $C\subm{e}$ divided by $T$. 
For the specific-heat data, its raw data before subtraction of the phononic part are presented in Extended Data Fig.~\ref{exfig_C_vs_T}.
Both susceptibility and specific heat exhibit a very sharp superconducting transition with $\Tc = 2.8$~K.
The susceptibility reaches the value for the full Meissner screening below around 2.5~K. 
Similarly, $C\subm{e}/T$ extrapolates almost to zero as $T \to 0$, suggesting nearly 100\%\ volume fraction. 
The sharp transition and almost perfect volume fraction again indicate high sample quality.

The observed specific-heat data are compared with a prediction of the standard Bardeen-Cooper-Schrieffer (BCS) theory.
We found noticeable difference from the prediction from the BCS theory (broken curve in Fig.~\ref{fig2}{\bf{b}}):
The observed jump $\Delta C\subm{e}/\Tc$ divided by the normal-state electronic specific heat coefficient $\gamma$ is 1.0, smaller than the BCS prediction ($\Delta C\subm{e}/\gamma \Tc = 1.43$). 
Moreover, $C\subm{e}/T$ below $\Tc$ is rather linear in $T$ in contrast to the exponential decay in the prediction.
This difference infers highly anisotropic and multi-band superconducting gap structure, which have been revealed in previous studies using various probes~\cite{Xu2021.PhysRevLett.127.187004,Duan2021.SciChinaPhysMechAstron.64.107462,Zhao2021.arxiv.2102.08356,Chen2021.Nature.599.222,Roppongi2022.arxiv.2206.02580}.

\section*{Quasi-two-dimensional superconductivity}

To study SC anisotropy, we used a vector magnet system with a horizontal rotation stage~\cite{Deguchi2004RSI}.
This system, the same one as used in Refs.~\cite{Yonezawa2017.NaturePhys.13.123}, allows us to perform accurate two-axis field-direction control.
The field is in-situ aligned to the sample by making use of the $\Hcc$ anisotropy, and the accuracy of the field alignment is better than 0.1 degrees (Extended Data Figs.~\ref{exfig.C-theta} and \ref{exfig.theta-peak}).
Throughout this paper, we define $\theta$ as the polar angle of the field measured from the $c$ axis and $\phi$ as the azimuth angle of the field along the $ab$ plane measured from the $a$ axis, as illustrated in Fig.~\ref{fig1}{\bf d}.

\begin{figure}
\begin{center}
\includegraphics[width=\linewidth]{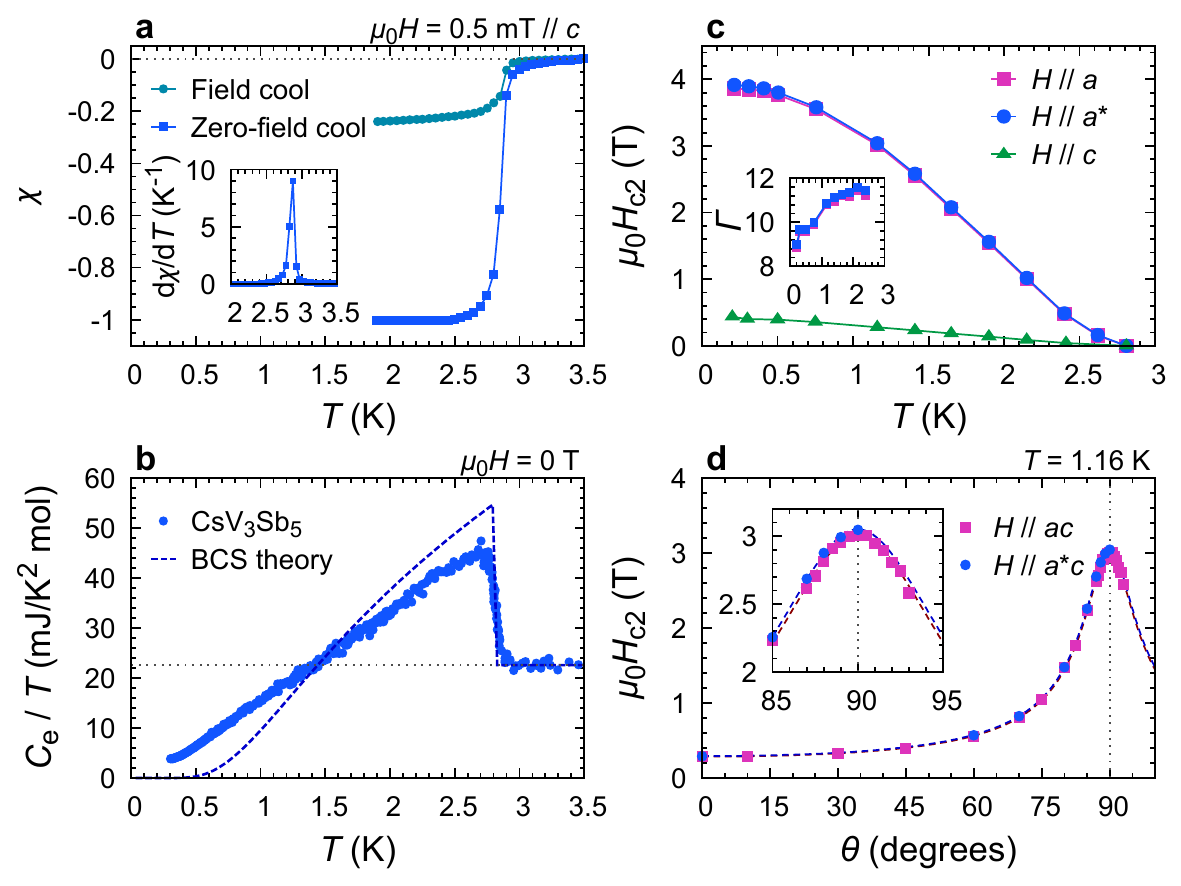}
\end{center}
\caption{
{\bf Zero-field superconducting properties and out-of-plane anisotropy of \csvs.}
{\bf a}. DC magnetic susceptibility $\chi$ of our \csvs\ single crystal measured under zero-field-cooled and field-cooled conditions. 
The demagnetization factor of 0.88 estimated from the sample shape has been considered. 
The inset shows the temperature derivative of zero-field-cooled $\chi$, showing a sharp superconductive transition at the critical temperature $\Tc =2.8$~K with a full-width of half-maximum of less than 0.1~K.
{\bf b}. Zero-field electronic specific heat of \csvs. 
The sharp jump at $\Tc$ again indicates high quality of the sample. The temperature dependence is clearly different from the prediction of the standard BCS theory (dotted curve). 
{\bf c}. Upper critical field $\Hcc$ along the $a$ (pink squares), $a^\ast$ (blue circles), and $c$ axes (green triangles). The inset shows the out-of-plane anisotropy $\Gamma \equiv H_{\mathrm{c2}\parallel a}/H_{\mathrm{c2}\parallel c}
$ (pink squares) and $H_{\mathrm{c2}\parallel a^\ast}/H_{\mathrm{c2}\parallel c}$ (blue circles). This quantity ranges from 9 to 11.5, indicative of Q2D nature of superconductivity.
{\bf d}. Polar field angle $\theta$ dependence of $\Hcc$ at 1.16~K in the $ac$ (pink squares) and $a^\ast c$ planes (blue circles). Both data sets are well fitted by using an anisotropic mass model indicated with the dotted curves. The fit yields the anisotropy parameter of 10.4 for the $ac$ plane and 10.5 for the $a^\ast c$ plane. The inset is an enlarged view near $\theta = 90\deg$.
\label{fig2}
}
\end{figure}

To evaluate the upper critical field $\Hcc$, we measured the field strength dependence of $C/T$ under various field directions (examplified in Extended Data Fig.~\ref{exfig_C_vs_H}). 
We then deduce $\Hcc$ from the anomaly in the $C(H)$ curves.
The obtained $\Hcc$ for three different crystalline directions, namely the $a$, $a^\ast$ and $c$ axes, are plotted in Fig.~\ref{fig2}{\bf c}.
As one can see, $\Hcc$ for the in-plane directions are quite similar but are nearly 10 times larger than that for the $c$ direction. 
For more quantitative comparison, we plot the out-of-plane anisotropy parameter $\Gamma \equiv H_{\mathrm{c2}\perp c} / H_{\mathrm{c2}\parallel c}$ in the inset of Fig.~\ref{fig2}{\bf c}. 
The anisotropy amounts to $11.4\pm 0.2$ above 1.6~K and reduces weakly to 8.9 at the lowest temperature. 
This high anisotropy indicates Q2D nature of superconductivity, as expected from high out-of-plane resistivity anisotropy ($\sim 600$)~\cite{Ortiz2020.PhysRevLett.125.247002} and from cylindrical Fermi surfaces~\cite{Neupert2022.NaturePhys.18.137} reflecting the layered crystal structure.

The out-of-plane anisotropy is studied in more detail from the polar angle $\theta$ dependence of $\Hcc$ shown in  Fig.~\ref{fig2}{\bf d}.
The $\Hcc$-$\theta$ curve measured at 1.16~K exhibits a sharp peak centered at $\theta = 90\deg$. 
The data is well fitted by the standard anisotropic mass model of the GL formalism: $\Hcc(\theta) = \Hcc(0\deg)/[\cos^2\theta + \Gamma^2\sin^2\theta]^{1/2}$ (dotted curves in the figure). 
The fit yields $\Gamma = 10.5$, in agreement with the $\Gamma$ values in the inset of Fig.~\ref{fig2}{\bf b}.
This successful fitting of the anisotropic mass model indicates that $\Hcc$ is dominated by the ordinary orbital pair-breaking effect and the GL theory well describes the out-of-plane anisotropy.
Moreover, the sharp peak in the $\Hcc(\theta)$ curve demonstrates that a very accurate field alignment is necessary to study intrinsic in-plane anisotropic properties since even a tiny out-of-plane misalignment can cause a sizable reduction of $\Hcc$.
In the present study, we emphasize that the field misalignment is carefully avoided as shown in the Extended Data Figs.~\ref{exfig.C-theta} and \ref{exfig.theta-peak}, and any extrinsic behavior originating field misalignment is absent (also see Method for quantitative discussion).

\section*{In-plane Hexagonal and Nematic Anisotropies}

\begin{figure}
\begin{center}
\includegraphics[width=\linewidth]{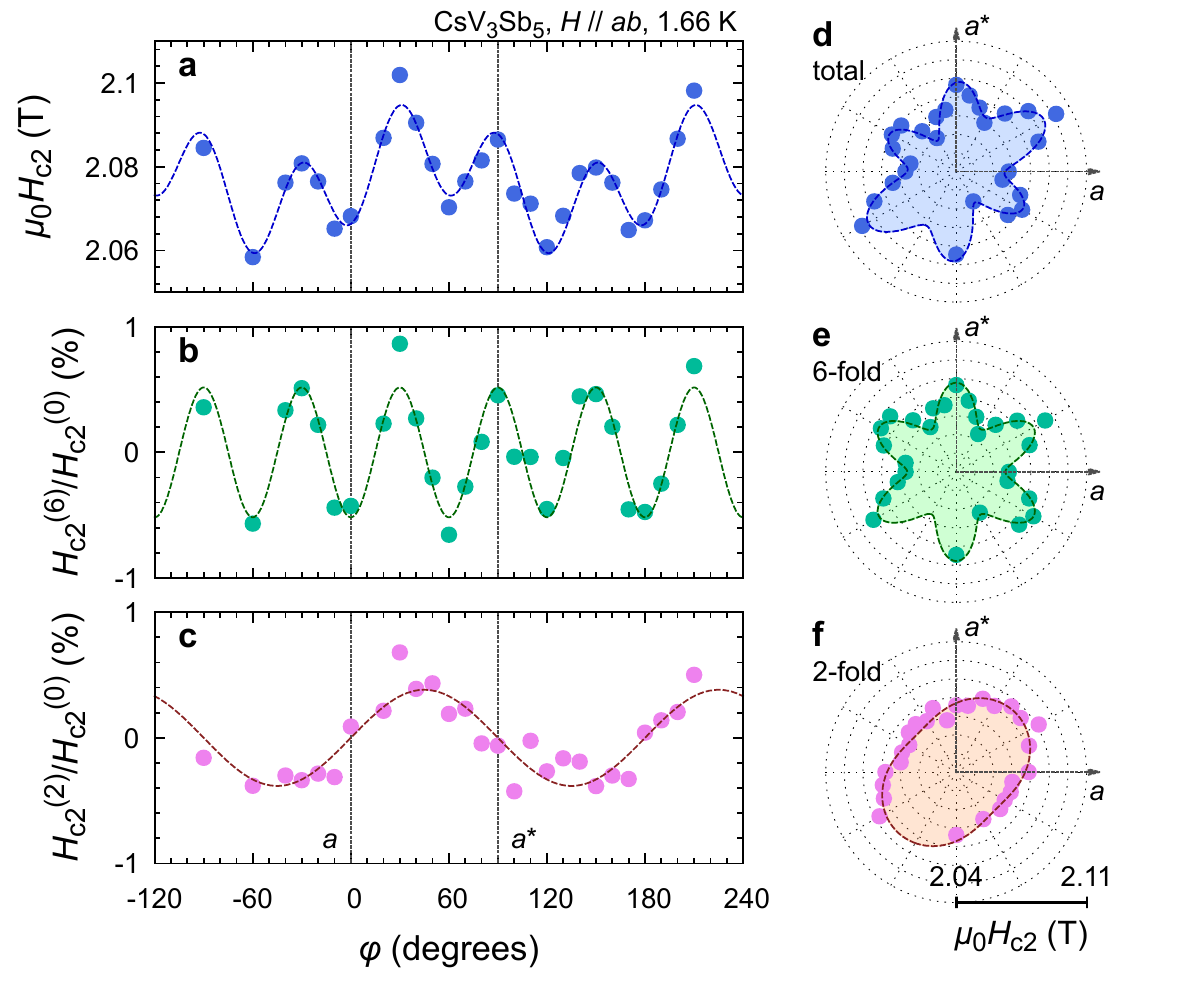}
\end{center}
\caption{
{\bf Unconventional in-plane anisotropy of superconductivity in \csvs.}
{\bf a}. In-plane field angle $\phi$ dependence of the upper critical field $\Hcc$. The dotted curve shows the result of fitting using a combination of six- and two-fold sinusoidal functions $\Hcc(\phi) = \Hsix\cos(6\phi) + \Htwo\sin(2\phi) + \Hzero$. 
{\bf b}. Relative six-fold component obtained by subtracting $\Htwo\sin(2\phi) + \Hzero$ from the data and by normalizing with $\Hzero$. 
{\bf c}. Relative two-fold anisotropic component of $\Hcc$ obtained by subtracting $\Hsix\sin(6\phi) + \Hzero$ from the data and subsequent normalization.
{\bf d-f}. Polar plots of the total $\Hcc$ ({\bf d}), and its six-fold ({\bf e}) and two-fold components ({\bf f}), illustrating coexistence of six-fold anisotropy with superconducting nematicity. 
\label{fig3}
}
\end{figure}

We now explain the novel in-plane anisotropy of $\Hcc$, which cannot be accounted for by the ordinary GL formalisms.
In Fig.~\ref{fig3}{\bf a}, we plot the in-plane field-angle $\phi$ dependence of $\Hcc$ measured at 1.66~K.
If the ERS is fully satisfied, no anisotropy would be visible.
However, surprisingly, the $\Hcc(\phi)$ data exhibits clear oscillation composed of six and two-fold components. 
Indeed, the data can be well fitted with the formula $\Hcc(\phi) = \Hsix\cos(6\phi) + \Htwo\sin(2\phi) + \Hzero$, as shown by the broken curve in Fig.~\ref{fig3}{\bf a}.
Here the parameters $\Hcc^{(i)}$ represent amplitudes of $i$-fold oscillatory components.
The obtained parameters are $\mu_0\Hsix = 0.0107 \pm 0.0008$~T, $\mu_0\Htwo = 0.0079 \pm 0.0009$~T, and $\mu_0\Hzero = 2.0770 \pm 0.0006$~T.
Figure~\ref{fig3}{\bf b} shows the relative six-fold component obtained by subtracting $\Htwo\sin(2\phi) + \Hzero$ from the data and normalized by $\Hzero$.
Similarly, the relative two-fold component shown in Fig.~\ref{fig3}{\bf c} was obtained by subtracting $\Hsix\sin(6\phi) + \Hzero$ from the data.
These data reveal that both anisotropy components are similar size and of the order of 0.5\% of $\Hzero$.
The raw $\Hcc$ data as well as the extracted components are also shown as polar plots in Figs.~\ref{fig3}{\bf d}-{\bf f}, in order to visually highlight the anisotropy behavior.

To evaluate the in-plane $\Hcc$ anisotropy in a more effective and sensitive way, we measured the specific heat in the middle of the SC transition. 
We fixed the magnetic field in the middle of the SC transition in the $C(H)/T$ curve (as illustrated in Extended Data Fig.~\ref{exfig_C-Hc2_schematics}), and then measure the field-angle dependence of the specific heat.
If there is a small change in $\Hcc$ as changing the field direction, we expect that $C(H)/T$ curves shift horizontally.
Approximating that this shift is a parallel shift, we can estimate the $\Hcc$ anisotropy from the measured specific-heat anisotropy $\delta C / T$ using the formula $\delta \Hcc \simeq -\alpha^{-1} (\delta C / T)$, where $\alpha = (1/T)(\mathrm{d}C /\mathrm{d}H)$ is the slope of $C/T$ vs $H$ curves at the SC transition.
We also comment that $C(\phi)/T$ in a field above $\Hcc$ does not show any detectable anisotropy (Extended Data Fig.~\ref{exfig_dHc2_vs_T_normal}), indicating that the measured $\delta C/T$ is free from anisotropies of any other origins (such as background or normal-state contributions).

\begin{figure}
\begin{center}
\includegraphics[width=12cm]{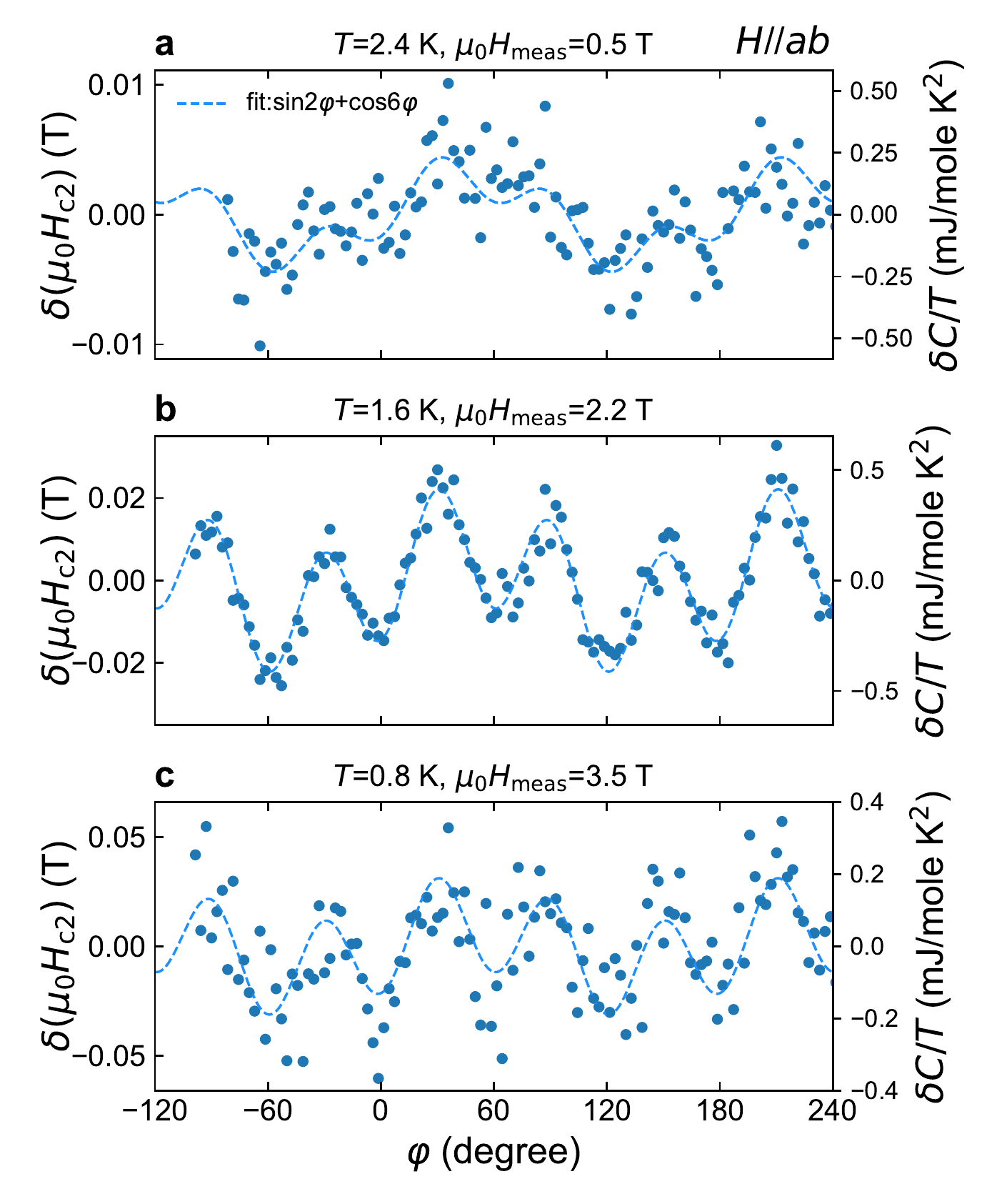}
\end{center}
\caption{
{\bf Temperature evolution of in-plane superconducting anisotropy in \csvs.} 
Here, in-plane anisotropy of $H_{\rm c2}$ deduced from the field-angle dependence of the specific heat at various temperatures are shown.
The measurement field $H\subm{meas}$ is close to the mid-point of the SC transition at each temperature.
{\bf a.} In-plane $\Hcc$ anisotropy at 2.4~K. As explained in the text, the anisotropy $\delta (\mu_0 \Hcc)$ is evaluated from the specific-heat anisotropy $\delta C/T$ using the relation $\delta (\mu_0 \Hcc) = -\alpha^{-1} \delta C/T$, where $\alpha = (1/T)(dC/dH)$ is the slope of the $C(H)/T$ vs. $H$ curve at the same temperature. The corresponding $\delta C/T$ values are shown using the right vertical axis. The curves represent the result of sinusoidal fitting using six and two fold oscillations.
{\bf b.} Same plot but for the data at 1.6~K.
{\bf c.} Same plot but for the data at 0.8~K.
\label{fig4}
}
\end{figure}

In Fig.~\ref{fig4}{\bf a}-{\bf c}, we plot the $\Hcc$ anisotropy obtained by this method at representative temperatures.
As one can see, the data set exhibits a combined six and two-fold oscillation, which is very similar to the data obtained from the direct $\Hcc$ measurement shown in Fig.~\ref{fig3}.
Thus these data were successfully fitted with the sinusoidal function $\delta\Hcc(\phi) = \Hsix\cos(6\phi) + \Htwo\sin(2\phi) $ as shown with the solid curves, and the anisotropy terms $\Hsix$ and $\Htwo$ are extracted at each temperature.

\begin{figure}
\begin{center}
\includegraphics[width=12cm]{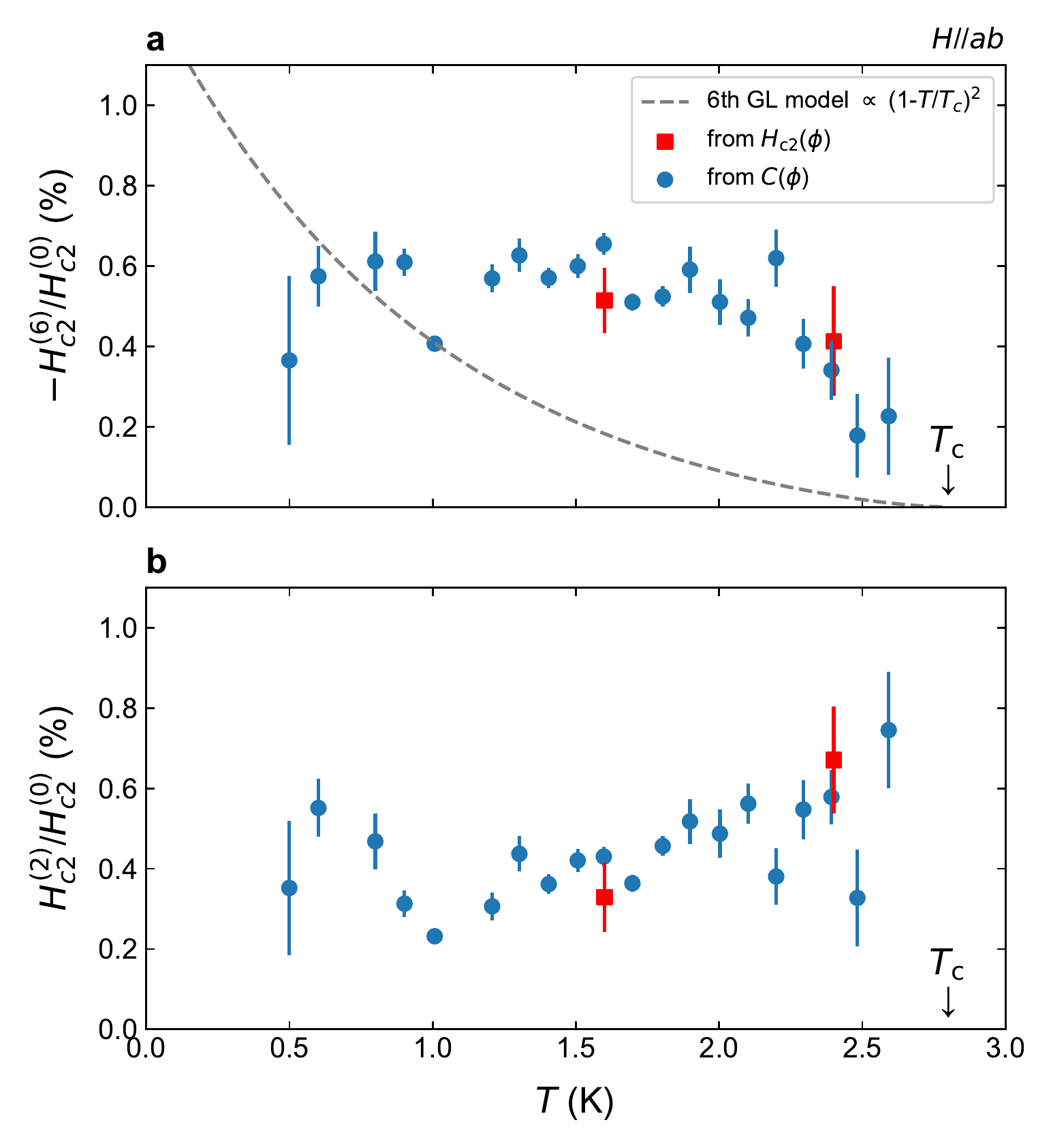}
\end{center}
\caption{
{\bf Temperature dependence of in-plane hexagonal and nematic $\Hcc$ anisotropy.}
{\bf a.} Temperature dependence of the six-fold hexagonal anisotropy component $\Hsix$ of $\Hcc$ divided by $\Hzero$, the averaged value of the in-plane $\Hcc$. 
Note that $-1$ is multiplied so that the ratio becomes positive.
The blue circles are obtained from the $C$-$\phi$ measurements as in Fig.~\ref{fig4}, whereas the red squares are from $\Hcc$-$\phi$ measurements as in Fig.~\ref{fig3}.
The error bar represents asymptotic standard errors of sinusoidal fittings of each $\Hcc(\phi)$ data set.
Clearly, the data deviates from the expectation under preserved ERS from the ordinary GL theory, $\Hsix/\Hzero \propto (1-T/\Tc)^2$, which is plotted using a broken curve.
{\bf b.} Temperature dependence of the two-fold anisotropy component $\Htwo$ divided by $\Hzero$.
\label{fig5}
}
\end{figure}

The ratios $\Hsix/\Hzero$ and $\Htwo/\Hzero$ are rather important since they represent hexagonal and nematic SC anisotropies of the material, respectively.
Thus, we plot temperature dependence of these ratios in Figs.~\ref{fig5}{\bf a} and {\bf b}.
Data from direct $\Hcc$ measurements (red points) and those from $C(\phi)$ measurements (blue points) agree very well.
It is quite intriguing that the hexagonal ratio $\Hsix/\Hzero$ stays almost constant up to 2.3~K, followed by a small downturn toward $\Tc = 2.8$~K. 
This downturn may be related to the small change in the slope of $\Hcc(T)$ curve at around 2.3~K (Fig.~\ref{fig2}{\bf c}), which can be attributed to multi-band effects.
The overall temperature dependence is in clear contrast to the standard GL theories predicting the ERS in hexagonal superconductors:
As explained before, GL theories predict that $\Hsix$ should be very small and, if exists, behave as $\Hsix \propto (1-T/\Tc)^3$ as $T \to \Tc$~\cite{Sauls1996.PhysRevB.53.8543,Krotkov2002.PhysRevB.65.224506}.
Since $\Hzero$ in GL theories behaves as $\Hzero \propto 1-T/\Tc$, the ratio $\Hsix/\Hzero$ should behave $\propto (1-T/\Tc)^2$ as shown with the broken curve in Fig.~\ref{fig5}{\bf a}.
However, this prediction cannot explain the observed temperature dependence at all.
We further examine the temperature dependence above 2.3~K, where the data exhibits downturn and approaches zero for $T \to \Tc$. 
However, as shown in the log-log plot of $\Hsix/\Hzero$ vs. $1-T/\Tc$ (Extended Data Fig.~\ref{exfig_H6-log}), the temperature dependence above 2.3~K is close to linear and far from quadratic.
Thus, this downturn is again hardly explained by the quadratic behavior predicted by the GL theory.
This data, to our knowledge, provide the first thermodynamic evidence for ERS violation in hexagonal superconductors.

The nematic anisotropy ratio $\Htwo/\Hzero$ is also quite important and interesting.
Firstly, this ratio remains finite even for thermodynamic measurements performed under the absence of field misalignment.
This fact indicates that the nematicity in the SC state of \csvs\ is intrinsic.
Nevertheless, we should stress that this intrinsic in-plane anisotropy is as small as a fraction of a percent.
This provides a clear demonstration that accurate field alignment is crucially important to evaluate intrinsic nematicity in materials with strong two-dimensionality.
Secondly, the temperature dependence of $\Htwo/\Hzero$ does not reach zero for $T \to \Tc$.
This feature suggests that the SC nematicity is inherited from the nematicity already existing outside of the SC phase~\cite{Teichler1975.PhysStatSolidiB.72.211}.
Indeed, various studies have revealed two-fold rotational behavior in electronic properties in the CDW state of \csvs~\cite{Neupert2022.NaturePhys.18.137,Xiang2021.NatureCommun.12.6727,Nie2022.Nature.604.59}. 
In that case, nematicity observed here is not a spontaneous feature of superconductivity.

\section*{Origins of the Unconventional Anisotropies}

Uncovering origins of violation of the ERS, namely robust six-fold in-plane $\Hcc$, in hexagonal superconductors has been big challenges in the past.
So far, as summarized in Table~\ref{tab.emergent-RS}, the only established explanation is a two-component SC parameter coupled with certain symmetry lowerings~\cite{Agterberg1995.PhysRevB.51.8481,Sauls1996.PhysRevB.53.8543,Krotkov2002.PhysRevB.65.224506,Venderbos2016.PhysRevB.94.094522}. 
In the context of the hexagonal heavy Fermion superconductor UPt\subs{3}, where a two-component SC order parameter ($E_{2u}$ or $E_{1u}$ states) is believed to be realized~\cite{Joynt2002.RevModPhys.74.235,Izawa2014.JPhysSocJpn.83.061013}, the six-fold $\Hcc$ anisotropy observed by magnetotransport~\cite{Keller1994.PhysRevLett.73.2364} has been attributed to the two-component superconductivity coupled to uniaxial symmetry breaking (weak antiferromagnetic order)~\cite{Agterberg1995.PhysRevB.51.8481,Sauls1996.PhysRevB.53.8543}, or to trigonal crystal distortion~\cite{Krotkov2002.PhysRevB.65.224506}. 
We comment that, for the former scenario, the direction of the uniaxial symmetry breaking is assumed to follow the external field direction due to the magneto-anisotropy of the weak antiferromagnetism, and the six-fold SC anisotropy appears through coupling between superconductivity and antiferromagnetism, whose free energy changes by the field direction~\cite{Agterberg1995.PhysRevB.51.8481,Sauls1996.PhysRevB.53.8543}.
These theories predict violation of the ERS and the hexagonal $\Hcc$ anisotropy to vary as $\propto 1-T/\Tc$.
Similar discussion has been developed more recently for doped \bs, where two-component nematic superconductivity is believed to be realized in a putative trigonal crystal lattice~\cite{Venderbos2016.PhysRevB.94.094522}. 

The present observation of six-fold $\Hcc$ in \csvs\ can be explained by similar scenario, namely two-component superconductivity coupled with underlying orthorhombic and/or trigonal symmetry breakings. 
Theoretically, possibility of two-component $d \pm id$ pairings have been discussed based on a model focusing on sublattice interference near the van-Hove singularities in the Kagome lattice~\cite{Yu2012.PhysRevB.85.144402,Wu2021.PhysRevLett.127.177001}.
A recent theory based on bond-order fluctuations also suggest possibility of $p_x$ and $p_y$ wave pairings in the clean limit~\cite{Tazai2022.SciAdv.8.eabl4108}, which also belong to two-component order parameters.
Moreover, in \csvs, the CDW order is known to exhibit electronic nematicity~\cite{Neupert2022.NaturePhys.18.137,Xiang2021.NatureCommun.12.6727,Nie2022.Nature.604.59}.
If such uniaxial symmetry breaking is rotatable by magnetic field direction, it can be one ingredient of the emergence of the clear six-fold anisotropy~\cite{Agterberg1995.PhysRevB.51.8481,Sauls1996.PhysRevB.53.8543}.
As another possibility, the lattice symmetry of the CDW phase is supposed to be trigonal, since the trigonal $P\bar{3}$ model provides the best fit of diffraction data~\cite{Ortiz2021.PhysRevX.11.041030}.
Such trigonal symmetry, together with two-component SC order parameter, can also explain the hexagonal SC anisotropy as proposed in Refs.~\cite{Krotkov2002.PhysRevB.65.224506,Venderbos2016.PhysRevB.94.094522}.

We should emphasize that this scenario of two-component superconductivity naturally explains the observed nematic anisotropy as well. 
When two-component order parameter $(\eta_1, \eta_2)$ is realized, the components $\eta_1$ and $\eta_2$ should be degenerate as long as the hexagonal or trigonal symmetry is preserved. 
In such cases, nematic anisotropy should be absent.
However, under the presence of uniaxial anisotropy in the CDW state as observed with many probes~\cite{Xiang2021.NatureCommun.12.6727,Nie2022.Nature.604.59}, the degeneracy between $\eta_1$ and $\eta_2$ must be lifted and can cause two-fold nematic anisotropy in $\Hcc(\phi)$.
This is quite similar to the situation in the \bs-based nematic superconductors~\cite{Yonezawa2017.NaturePhys.13.123,Yonezawa2019.condmat.4.2}, where certain symmetry-breaking field is believed to cause large two-fold anisotropy in SC quantities.
In contrast, if an ordinary single-component SC state is realized, the two-fold $\Hcc$ anisotropy should occur through the anisotropy of the Fermi velocity $\vF$; i.e. $H_{\mathrm{c2}\parallel a} / H_{\mathrm{c2}\parallel a^\ast} = v\subm{F}^{a} / v\subm{F}^{a^\ast}$. 
However, the Fermi velocity anisotropy in the CDW state seems to be very small: cAngle-resolved photoemission spectroscopy (ARPES) of \csvs~\cite{Nakayama2021.PhysRevB.104.L161112,Kang2022.NaturePhys.18.301} does not resolve nematicity in the electronic bands and the resistivity anisotropy observed in elasto-resistivity is of the order of 0.1\% under 100~ppm strain~\cite{Nie2022.Nature.604.59}, much smaller than those observed in iron-based superconductors~\cite{Chu2012.Science.337.6095}.
Thus, the nematic $\Hcc$ anisotropy is more difficult to be explained under the assumption of single-component superconductivity.

Nevertheless, we should comment here that some experiments support rather conventional SC order parameters~\cite{Mu2021.ChinPhysLett.38.077402,Duan2021.SciChinaPhysMechAstron.64.107462,Roppongi2022.arxiv.2206.02580}. 
If conventional pairing is truly realized, one needs to include other terms that are absent in previous GL theories~\cite{Burlachkov1985.SovPhysJETP.62.800,Agterberg1995.PhysRevB.51.8481,Sauls1996.PhysRevB.53.8543,Krotkov2002.PhysRevB.65.224506}, in order to explain the observed hexagonal anisotropy.
For example, by taking into account the multi-band nature of superconductivity, one may be able to explain hexagonal anisotropy.
As another exotic possibility, roton pair-density wave (PDW), which has recently been proposed~\cite{Chen2021.Nature.599.222}, may also explain the hexagonal anisotropy once its GL terms are included.
Thus, we cannot deny novel coupling between conventional superconductivity and underlying electronic properties of the Kagome metal may explain the observed hexagonal anisotropy.
But we would like to emphasize here that, regardless of the origin, the observation of ERS violation itself is very rare in any known hexagonal superconductors.

Our first thermodynamic evidence for violation of the emergent rotational symmetry together with nematicity clearly indicates that superconductivity in \csvs\ is exotic. 
Most plausible explanation using existing GL theories is two-component order parameter intertwined with CDW symmetry lowerings.
The present work should stimulate further studies to search for possible unconventional SC properties originating from this unique order parameter, such as time-reversal-symmetry breaking and strong response to uniaxial deformation.
Inclusion of novel ingredients, such as multi-band, CDW, and PDW contributions, may explain our observation even for single-component order parameter, which can be another direction of theoretical frontier in the Kagome metals \Avs.
Finally, our results shed light on special importance of hexagonal superconductors, which can host unique states with a help of high rotational symmetry and intertwined electronic orders.

\section*{Methods}

\subsection*{Sample preparation and characterization}

High-quality single crystals of \csvs\ were grown by a self-flux method with binary Cs-Sb as flux. 
The molar ratio of Cs$_{0.4}$Sb$_{0.6}$:CsV$_3$Sb$_5$ was chosen to be 20:1. 
The raw material was loaded in an alumina crucible, which was then sealed in an evacuated quartz tube with vacuum of $2\times 10^{-4}$~Pa. 
The tube was put into a muffle furnace and heated to 1000$^\circ$C at a rate of 5$^\circ$C/h. 
After maintaining at 1000$^\circ$C for 12~h, 
the tube was cooled to 200$^\circ$Cat a rate of 3$^\circ$C/h and subsequently down to room temperature with the furnace switched off. The flux was removed by distilled water, and finally shiny crystals with hexagonal shape were obtained.

After the growth, each crystal was examined by means of the Laue photographs and magnetization measurements.
Laue photographs were taken by using a commercial Laue camera (Rigaku, RASCO‐BL2) with a backscattering geometry, with incident X-ray along the $c$ axis. 
As exemplified in Fig.~\ref{fig1}{\bf f}, high-quality crystals exhibit clear Laue spots. 
The DC magnetization was measured using a commercial magnetometer equipped with superconducting quantum interference device (SQUID) detection coils (Quantum Design, MPMS-XL).
A sample was mounted in a plastic straw carefully so that we do not apply excessive strain that can degrade crystallinity.
A crystal exhibiting sharp SC transition with nearly 100\% volume fraction (Fig.~\ref{fig2}{\bf a}, {\bf b}) was chosen and used in the following specific-heat measurements.

\subsection*{Calorimetry}

The specific heat was measured using a home-made low-background calorimeter~\cite{Yonezawa2017.NaturePhys.13.123}. 
The sample was sandwiched by a heater and thermometer, for which commercial ruthenium-oxide resistance chips were used.
The background contribution was measured using a copper plate as a reference sample and was subtracted from the data.
We employed the AC method to measure the heat capacity.
We typically chose frequency of the AC heater current $\omega\subm{H}/2$ to be 0.1-0.5~Hz depending on the condition to maximize the sensitivity and accuracy. 
The oscillatory component of the temperature $T\subm{AC}$ and its phase shift $\delta$ was measured by using lock-in amplifiers (Stanford Research Systems, SR830). 
Then the heat capacity $C$ is evaluated by the relation~\cite{Velichkov1992.Cryogenics.32.285}
\[
C = \frac{P_0}{2\omega\subm{H}T\subm{AC}} \sin\delta ,
\]
where $P_0$ is the power produced by the heater.
We adjusted $P_0$ so that $T\subm{AC}$ is a few percent of the sample temperature.

The calorimeter was placed in a ${}^3$He-${}^4$He dilution refrigerator (Oxford Instruments, Kelvinox 25) and cooled down to around 0.2~K.

\subsection*{Magnetic field control}

The magnetic field was applied using the vector magnet system~\cite{Deguchi2004RSI}.
This system consists of two orthogonal SC magnets (vertical $z$ solenoid magnet up to 3~T  and horizontal $x$ split magnet up to 5~T; Cryomagnetics, VSC-3050) placed on a horizontal rotating stage. 
By controlling the relative strengths of the magnet currents to the $x$ and $z$ magnets, we can rotate the magnetic field vertically in the laboratory frame, and by rotating the stage, we can control the horizontal field direction. 

The crystalline directions of the sample placed in the vector magnet was determined by making use of the $\Hcc$ anisotropy. The $ab$ plane was determined by detecting the SC response in the specific heat while rotating the magnetic field in the laboratory frame with field strength close to the in-plane $\Hcc$.
As explained in Extended Data Figs.~\ref{exfig.C-theta} and \ref{exfig.theta-peak}, the accuracy between the determined and actual sample frames are better than 0.1$^\circ$ in the $\theta$ direction. 
Then, within the determined $ab$ plane, the directions of the $a$ and $a^\star$ axes were determined based on the six-fold $\Hcc$ oscillation with a help of Laue photos taken before the cooling.
The accuracy in the in-plane angle $\phi$ is around a few degree.
After these processes, we can obtain a conversion matrix between the laboratory and sample frames.
The field angles shown in this Article is all expressed in the sample frame. 

As explained above, the out-of-plane field misalignment is less than 0.1$^\circ$ in $\theta$ direction. 
Assuming the anisotropic effective mass model determined from the data in Fig.~\ref{fig2}{\bf d}, misalignment of 0.1$^\circ$, if exists, would results in extrinsic two-fold $\Hcc$ anisotropy of 0.0005~T at 1.16~K, which is as small as 0.016\% of the in-plane $\Hcc$ of 3.05~T.
This is much smaller than the observed nematic anisotropy (0.2-0.6\%; see Fig.~\ref{fig5}{\bf b}). 
Therefore, the extrinsic two-fold anisotropy originating from field misalignment is totally negligible in our study.

\section*{Acknowledgments}

We acknowledge Y.~Yanase, G.~Mattoni, S.~Kitagawa, K.~Ishida, J.~W.~F.~Venderbos for fruitful discussion.
The work at Kyoto Univ. was supported by Grant-in-Aids for 
Scientific Research on Innovative Areas ``Quantum Liquid Crystals'' (KAKENHI Grant Nos. 20H05158, 22H04473) from the Japan Society for the Promotion of Science (JSPS),
a Grant-in-Aid for JSPS Fellows (KAKENHI Grant No. 20F20020) from JSPS,
Grant-in-Aids for Scientific Research (KAKENHI Grant Nos. 17H06136, 22H01168) from JSPS,
by Core-to-Core Program (No. JPJSCCA20170002) from JSPS
by a research support funding from The Kyoto University Foundation, 
and by ISHIZUE 2020 of Kyoto University Research Development Program.
The work at Beijing Institute of Technology was supported by the Natural Science Foundation of China (Grant No.~92065109), the National Key R\&D Program of China (Grant Nos.~2020YFA0308800, 2022YFA1403400), and the Beijing Natural Science Foundation (Grant Nos.~Z210006, Z190006). 

\section*{Author contributions}

Z.~Wang and S.~Yonezawa designed this study.
K.~Fukushima, K.~Obata, and S.~Yonezawa performed specific-heat measurements and analyses with guidance of Y.~Maeno. 
Z.~Wang, Y.~Li, and Y.~Yao grew single crystalline samples.
K.~Fukushima, K.~Obata, S.~Yamane, and Y.~Hu characterized samples with guidance of  Y.~Maeno and S.~Yonezawa.
The manuscript was prepared mainly by S.~Yonezawa and K.~Fukushima based on discussion among all authors.

\section*{Competing financial interests}

All authors declare there is no competing interests regarding this work.


\clearpage

\section*{Figures}

\clearpage

\clearpage

\clearpage
\clearpage
\clearpage

\renewcommand{\arraystretch}{1.5}

\begin{table}[h]
\begin{center}
\caption{ {\bf Violation of the emergent rotational symmetry in hexagonal superconductors.} Predictions based on GL theories are summarized for cases of single- and two-component SC order parameters with or without weak symmetry lowering from perfect hexagonal lattice.
When the emergent rotational symmetry is preserved, the hexagonal $\Hcc$ anisotropy component, $\Hsix$, should be small and behave as $\propto (1-T/T_{c})^{3}$. In contrast, it should behave as $\propto (1-T/T_{c})^{1}$ if the emergent rotational symmetry is violated, as observed in the present experiment in \csvs.}
\label{tab.emergent-RS}
\begin{tabular}{p{0.2\textwidth}p{0.2\textwidth}p{0.3\textwidth}c}
\hline
\centering SC order parameter & \centering Symmetry lowering & \centering Emergent rotational symmetry  & Refs. \\
\hline\hline 
\centering Single-component & \centering None & \centering Preserved & \cite{Sauls1996.PhysRevB.53.8543,Krotkov2002.PhysRevB.65.224506} \\
\centering Single-component & \centering Orthorhombic & \centering Preserved  & \cite{Teichler1975.PhysStatSolidiB.72.211} \\
\centering Single-component & \centering Trigonal & \centering Preserved & \cite{Venderbos2016.PhysRevB.94.094522} \\
\hline
\centering Two-component & \centering None & \centering Preserved  & \cite{Burlachkov1985.SovPhysJETP.62.800,Agterberg1995.PhysRevB.51.8481,Sauls1996.PhysRevB.53.8543,Krotkov2002.PhysRevB.65.224506} \\
\centering Two-component & \centering Orthorhombic$^\ast$ & \centering {\bf Violated}  & \cite{Agterberg1995.PhysRevB.51.8481,Sauls1996.PhysRevB.53.8543} \\
\centering Two-component & \centering  Trigonal &\centering  {\bf Violated}   & \cite{Krotkov2002.PhysRevB.65.224506,Venderbos2016.PhysRevB.94.094522} \\ 
\hline
\end{tabular}
\end{center}

\vspace{1em}
 
{\ \ \ \ $^\ast$ Direction of the symmetry breaking field must follow the magnetic field direction}

\end{table}

\clearpage

\setcounter{figure}{0}
\renewcommand{\figurename}{Extended Data Fig.}

\section*{Extended Data}

\begin{figure}[h]
\begin{center}
\includegraphics[width=10cm]{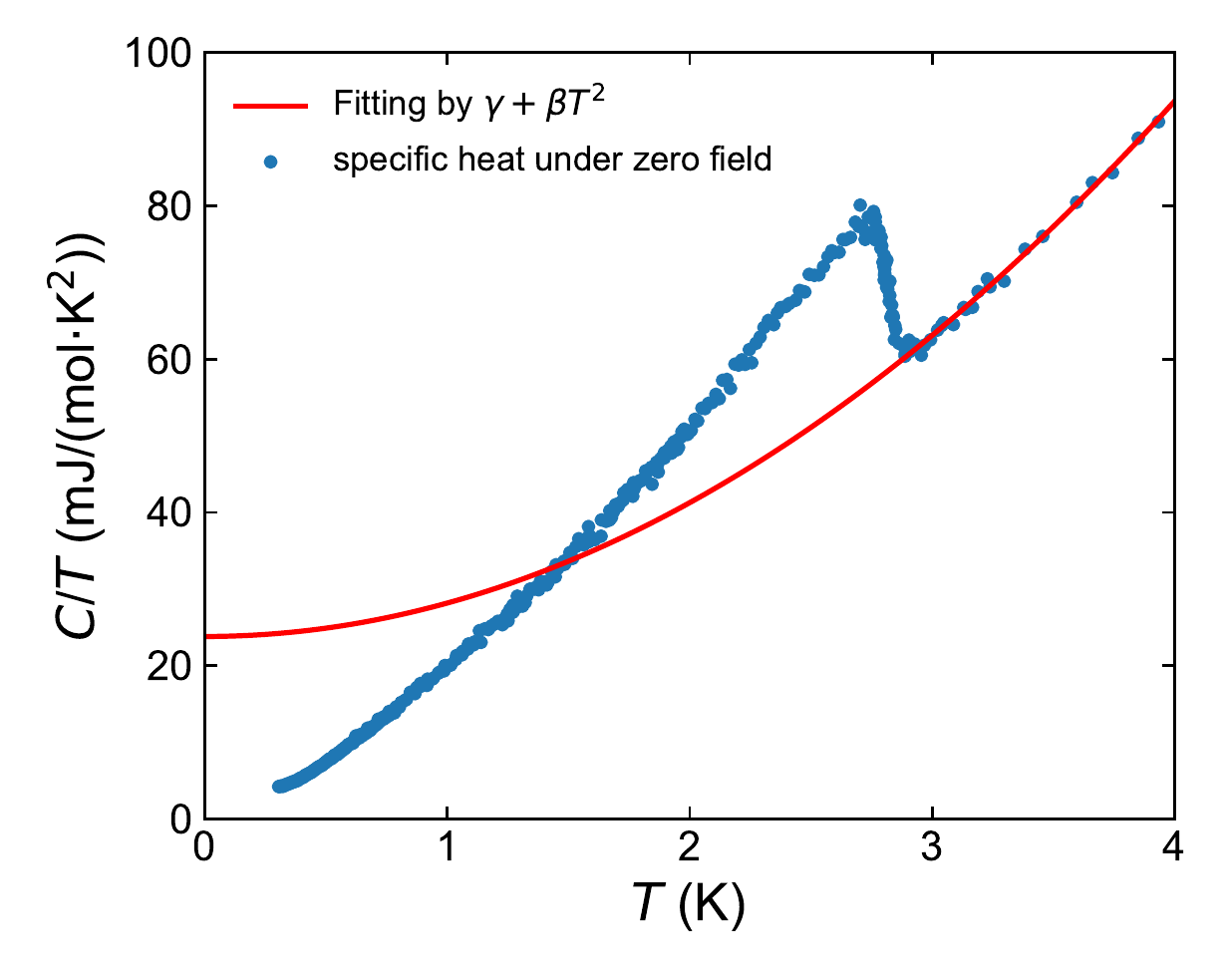}
\end{center}
\caption{
{\bf Temperature dependence of the specific heat under zero field.}
Background (addenda) contribution is removed. 
The red curve is the result of fitting of the function $C/T = \gamma + \beta T^2$ to the normal-state data from 3.1 to 4.0~K.
This fitting yields $\gamma = 23.81$~mJ/K$^2$mol and $\beta = 4.364$~mJ/K$^4$mol, respectively.
\label{exfig_C_vs_T}
}
\end{figure}

\clearpage

\begin{figure}[p]
\begin{center}
\includegraphics[width=9cm]{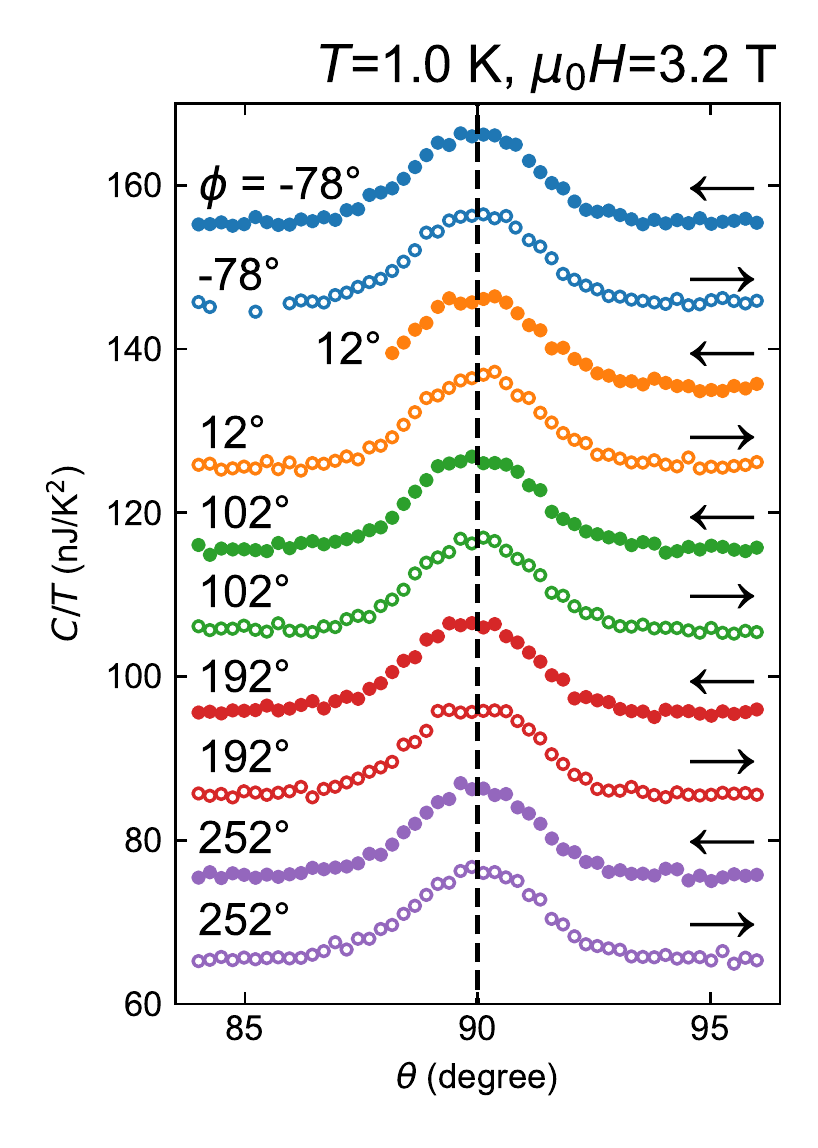}
\end{center}
\caption{
{\bf Polar-angle dependence of specific heat at several azimuth angles.}
Each curve is offset for clarity.
The arrows means the direction of the measurement sweeps.
In these plots, the specific-heat peak appears when the field is exactly parallel to the $ab$ plane. 
After the conversion matrix between the laboratory and sample frames are determined, this data was measured in order to check the accuracy of the determined coordinate conversion. 
As one can see, for all the curves, the peak appears at  $\theta=90^{\circ}$ indicated by the black vertical line.
Moreover, all $C(\theta)/T$ curves are symmetric with respect to $\theta=90^{\circ}$. 
These facts manifest that the determined sample frame exactly matches with the actual sample crystalline axes. 
Thus, in the data presented in this paper, field misalignment is negligibly small.
\label{exfig.C-theta}
}
\end{figure}

\begin{figure}[p]
\begin{center}
\includegraphics[width=12cm]{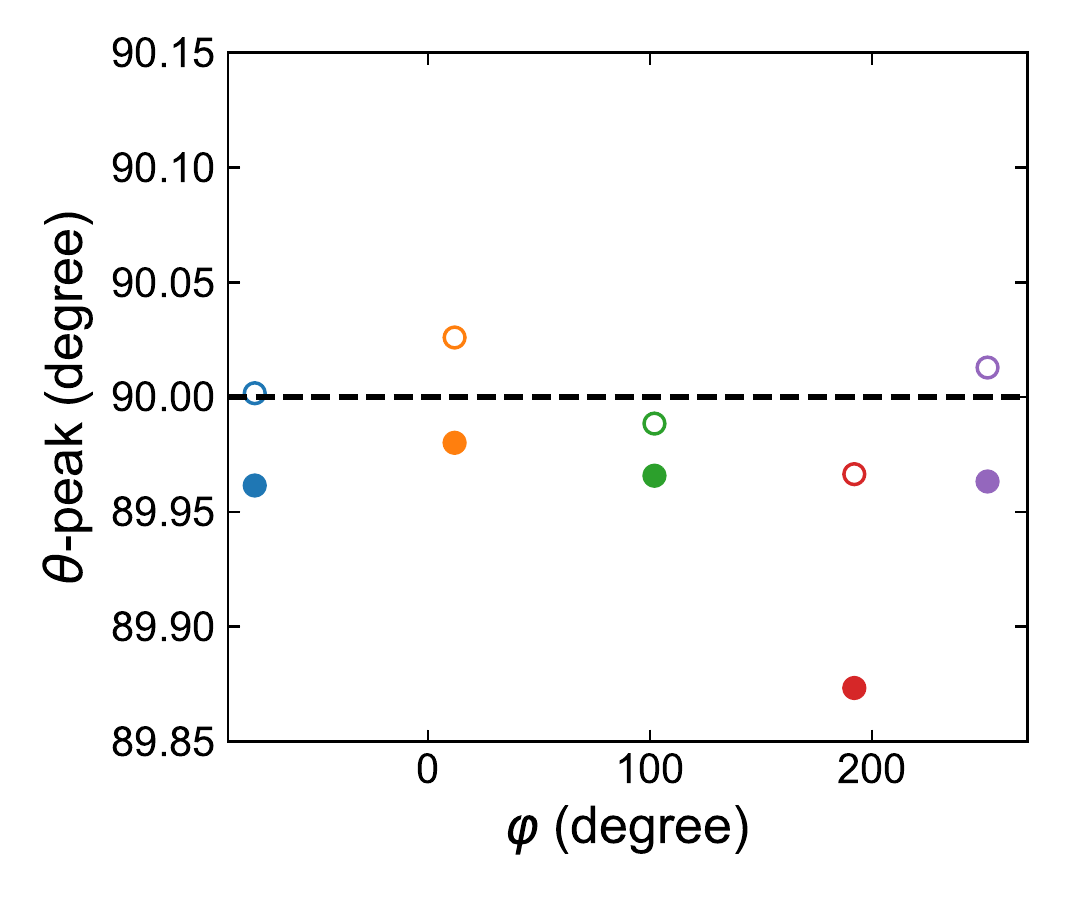}
\end{center}
\caption{
{\bf Accuracy of the magnetic-field alignment.}
Here, the location of the peak of $C(\theta)/T$ curves in Extended Data Fig.~\ref{exfig.C-theta} as a function of the azimuth angle.
The peak positions of the $C(\theta)/T$ curves are obtained by fitted the curves with a fourth-order polynomial with even order terms, and plotted against the azimuth angle. 
The colors in this plot correspond to the colors in Extended Data Fig.~\ref{exfig.C-theta}, and the open and closed symbols are from up and down sweeps, respectively.
The horizontal dotted line marks $\theta=90^{\circ}$. 
Fitting this data with a sin curve gives the amplitude of about 0.04$\deg$, which provides the upper limit of the alignment accuracy. 
\label{exfig.theta-peak}
}
\end{figure}

\begin{figure}[p]
\begin{center}
\includegraphics[width=12cm]{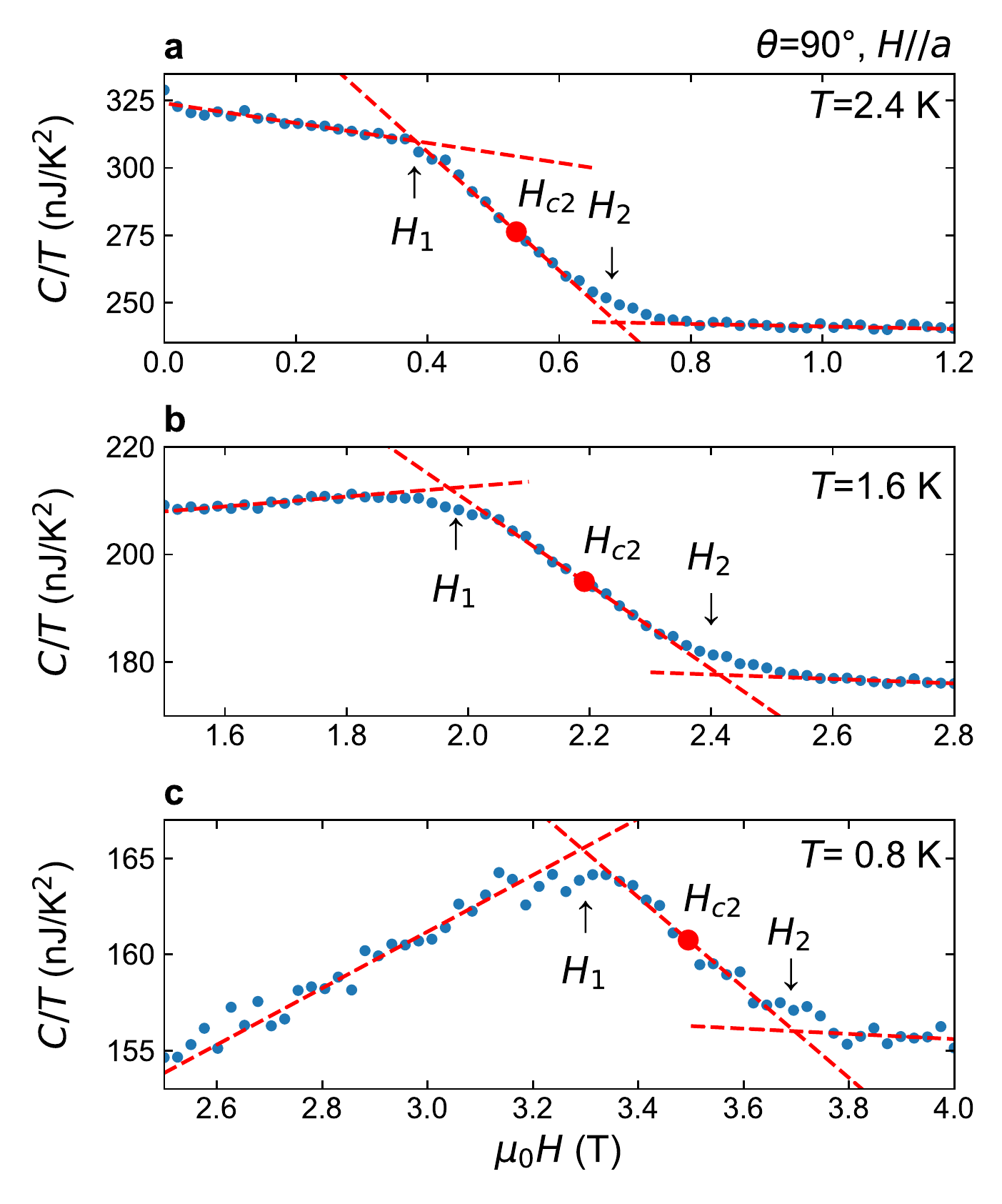}
\caption{
{\bf Representative magnetic-field strength dependence of the heat capacity near $\Hcc$.}
{\bf a.} Magnetic field dependence of the heat capacity at 2.4~K and $\phi = 0\deg$. 
The red broken lines are results of linear fittings of the $C(H)/T$ curves to evaluate $\Hcc$.
The intersections of the fit results are defined as $H_1$ and $H_2$ as illustrated in the figure.
Then, $\Hcc$ is defined as $\Hcc=(H_1+H_2)/2$.
{\bf b} and {\bf c}. Same plots but for 1.6~K and 0.8~K.
\label{exfig_C_vs_H}
}
\end{center}
\end{figure}

\begin{figure}[p]
\begin{center}
\includegraphics[width=12cm]{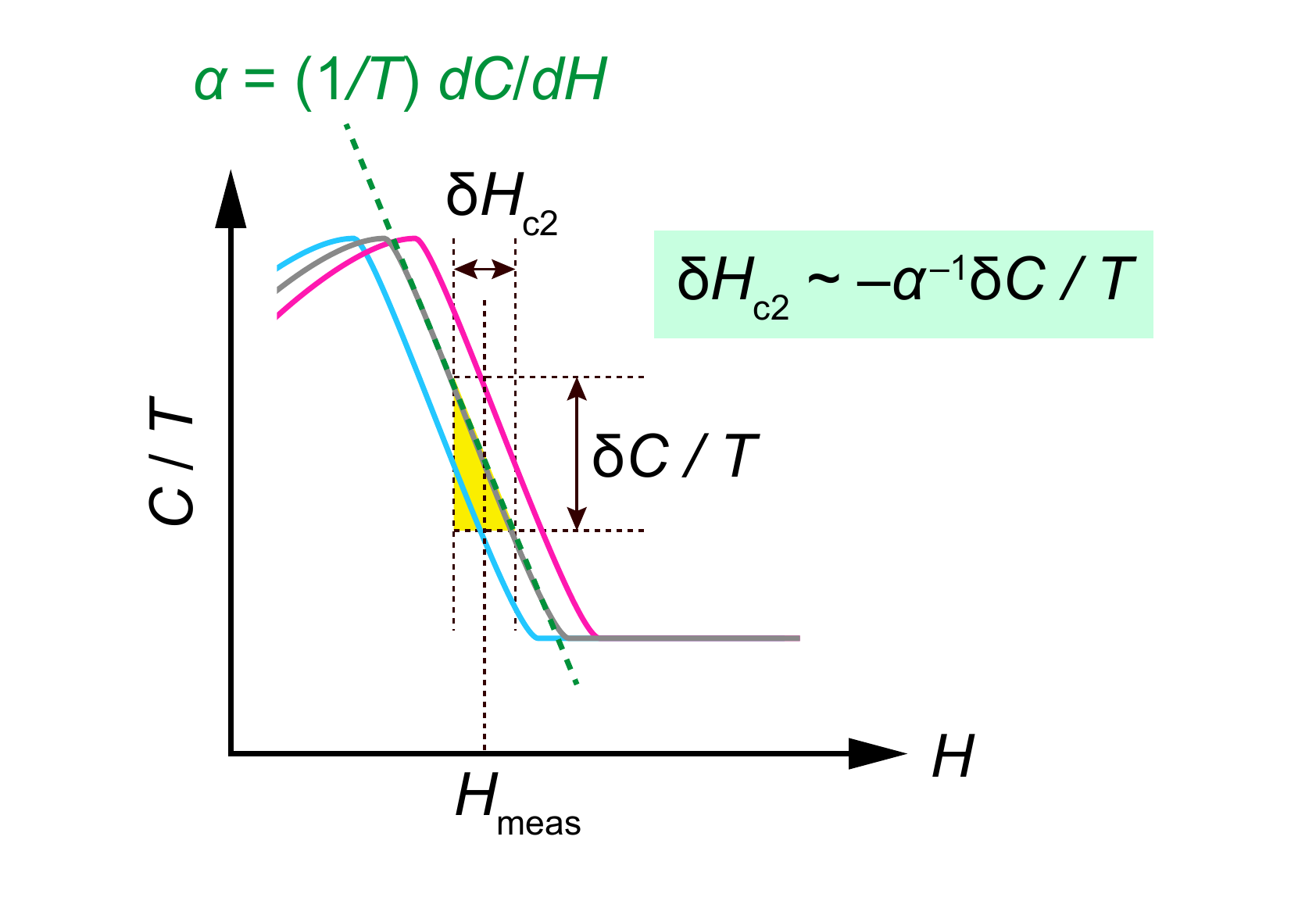}
\end{center}
\caption{
{\bf Schematic description of $\Hcc$ anisotropy evaluation from in-plane specific heat anisotropy.}
Here, the field strength dependence of $C(H)/T$ has a slope $\alpha = (1/T) dC/dH$ around the upper critical field $\Hcc$. 
When $\Hcc$ changes by a small value $\delta \Hcc$, the $C(H)/T$ curve exhibits a horizontal shift as depicted by the thick light-blue and pink curves.
Thus, if we measure $C/T$ at a fixed field $H\subm{meas}$ close to the mid-point of the SC transition, the specific heat changes by $\delta C/T$.
With simple geographic consideration, $\delta \Hcc$ and $\delta C/T$ are related through the formula $ (\delta C/T) / (\delta \Hcc) \simeq -\alpha$, leading to $\delta \Hcc \simeq -\alpha^{-1} \delta C/T$.
\label{exfig_C-Hc2_schematics}
}
\end{figure}

\clearpage

\begin{figure}[p]
\begin{center}
\includegraphics[width=10cm]{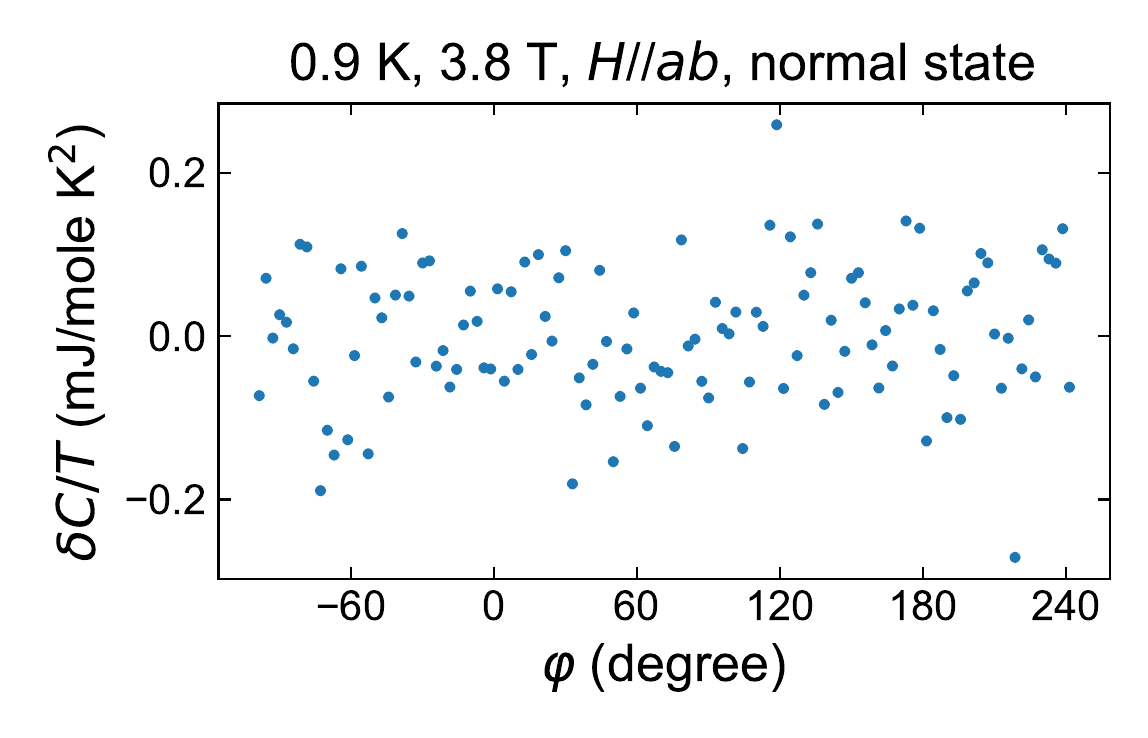}
\end{center}
\caption{
{\bf  Normal-state in-plane azimuth angle $\varphi$ dependence of the specific heat  of \csvs.}
Here, the specific heat was measured at 0.9~K under 3.8~T, which is higher than $\Hcc$ at this temperature ($\mu_0\Hcc = 3.4$~T; see Fig.~\ref{fig2}{\bf c}). 
There is no detectable anisotropy in the normal-conducting CDW state above $\Hcc$. 
\label{exfig_dHc2_vs_T_normal}
}
\end{figure}

\clearpage

\begin{figure}[p]
\begin{center}
\includegraphics[width=10cm]{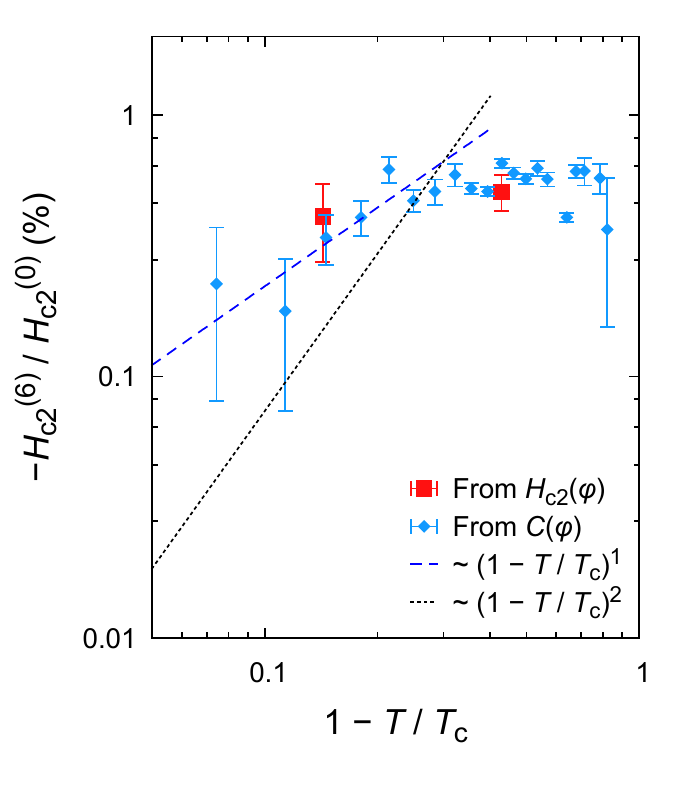}
\end{center}
\caption{
{\bf Log-log plot of $-\Hsix/\Hzero$ as a function of the reduced temperature $1-T/\Tc$.}
This graph is intended to investigate the temperature dependence of the hexagonal anisotropy $-\Hsix/\Hzero$ above 2.3~K ($1-T/\Tc \lesssim 0.18$), where this ratio exhibits downturn as shown in Fig.~\ref{fig5}{\bf a}.
This log-log plot show that $-\Hsix/\Hzero$ is linearly proportional to $1-T/\Tc$ (blue broken line), which is distinct from the quadratic behavior $\propto (1-T/\Tc)^2$ (black dotted line) predicted by standard GL theories.
\label{exfig_H6-log}
}
\end{figure}

\end{document}